\documentclass[]{nature}
\usepackage{graphicx}
\usepackage{amsmath}
\usepackage{amssymb}
\usepackage[english]{babel}
\begin{document}

\title{Coherent two-mode dynamics of a nanowire force sensor}

\author{Floris R. Braakman$^1$, Nicola Rossi$^1$, G\"ozde T\"ut\"unc\"uoglu$^2$, Anna
  Fontcuberta i Morral$^2$, \& Martino Poggio$^1$}
\maketitle

\begin{affiliations}
 \item Department of Physics, University of Basel, Klingelbergstrasse 82, 4056 Basel, Switzerland
 \item Laboratiore des Mat\'{e}riaux Semiconducteurs, Institut des Mat\'{e}riaux, \'{E}cole Polytechnique F\'{e}d\'{e}rale de Lausanne, 1015 Lausanne, Switzerland
\end{affiliations}

\begin{abstract}Classically coherent dynamics analogous to those of quantum two-level systems are studied in the setting of force sensing. We demonstrate quantitative control over the coupling between two orthogonal mechanical modes of a nanowire cantilever, through measurement of avoided crossings as we deterministically position the nanowire inside an electric field. Furthermore, we demonstrate Rabi oscillations between the two mechanical modes in the strong coupling regime. These results give prospects of implementing coherent two-mode control techniques for force sensing signal enhancement.
\end{abstract}

\section*{Introduction}
The coherent dynamics of two-level systems has been studied mainly in the regime of quantum physics\cite{CohenTannoudji04}. However, many concepts that are familiar in this context have counterparts in classical systems\cite{Dragoman04}. Although some classical analogues have already been studied\cite{Spreeuw90,Spreeuw93} several decades ago, interest has increased in recent years, particularly in the context of nanoscale mechanical resonators. Researchers have demonstrated strong coupling of two modes of mechanical oscillators\cite{Faust12,Faust13,Okamoto13}, driven Rabi oscillations\cite{Rabi37,Faust13,Okamoto13}, St\"uckelberg interferometry\cite{Seitner16,Seitner17}, and have measured coherence and dephasing times\cite{Faust13,Schneider14}. 

\noindent Recently, nanowire (NW) mechanical oscillators have been used for force\cite{Poggio13,Nichol12,Nichol13} and mass sensing\cite{Gil-Santos10,Vidal-Alvarez15}. NWs have favorable properties for sensing such as a cantilever geometry, low mass, low mechanical dissipation, and high oscillation frequencies. Moreover, several experiments have shown that NWs typically possess two nearly degenerate, orthogonal flexural modes\cite{Nichol08,Gloppe14,Cadeddu16}, which have been used to demonstrate two-dimensional lateral force sensing\cite{Gloppe14,Rossi17,Lepinay17}. Sensing based on such two-mode nanomechanical oscillators could be enhanced through the use of coherent two-mode dynamics. In particular, coherent pulse sequences reminiscent from quantum control techniques\cite{Degen17} give the potential to increase the frequency stability of mechanical sensors, ultimately leading to higher sensitivities. Finally, such pulsing techniques make it possible to implement a range of noise spectroscopy methods in classical force and mass sensing\cite{Degen17}.\\

\noindent In this article, we take the first step towards this goal by investigating the coherent dynamics of the two orthogonal fundamental flexural modes of a NW cantilever. First, we spatially image the two-dimensional lateral force gradients acting on the NW when it is placed inside an electric field created by gate electrodes patterned on a nearby sample surface. Next, we demonstrate strong coupling between the two modes through the measurement of avoided crossings of the mode frequencies. The spatial measurement of the force gradients allows for careful tuning of the mode coupling strength produced by the derivative of shear forces. This coupling is then used to implement Rabi oscillations between the populations of the two modes. 

\section*{Setup and measurement techniques}
\begin{figure*}
\includegraphics[width=1\textwidth]{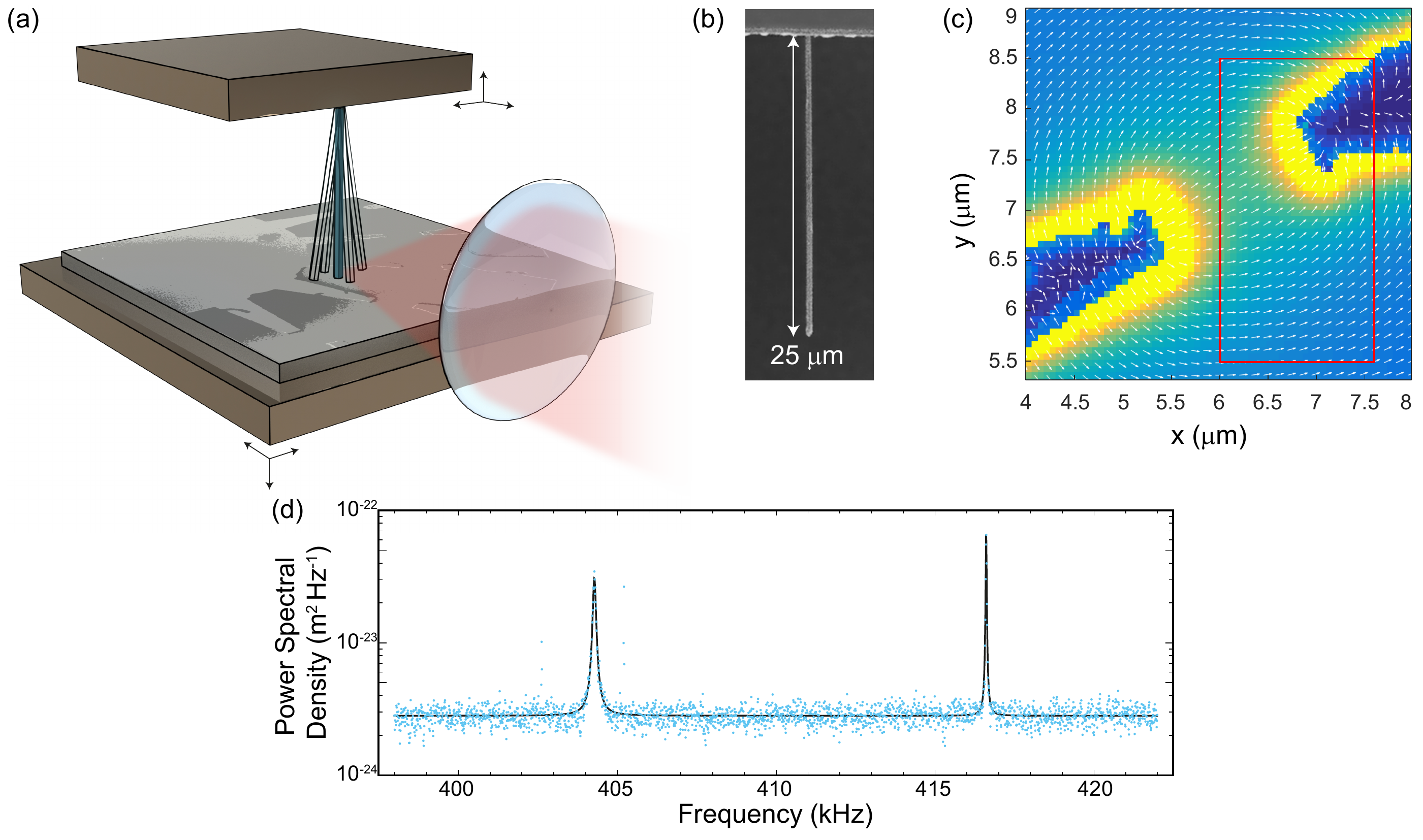}
\caption{(a) Overview of the experimental setup. (b) SEM of the NW used in this work. (c) Simulated electric field of the gate electrodes, with the tip of the NW placed at a height of 70 nm above the sample surface. The arrows indicate the direction of the field, while the magnitude is encoded in colorscale. The red rectangle indicates the region of the sample measured in the remainder of this work. (d) Interferometric measurement of the power spectral density of the two fundamental flexural modes of the NW, taken with a tip-sample distance of 70 nm. }
\label{fig:Figure1}
\end{figure*}

\noindent We perform vectorial scanning force microscopy using a single GaAs/AlGaAs NW cantilever. NWs were grown onto a Si (111) substrate through Molecular Beam Epitaxy using the Ga-assisted method\cite{Colombo08}. We removed excess NWs, such that individual NWs could be used for the experiments. The NW used to perform the measurements presented here, has a length of 25 $\mu$m and a diameter of 350 nm, as determined by scanning electron microscopy (SEM, see Fig.~\ref{fig:Figure1}b). Such NWs feature a hexagonal cross-section, which regularly exhibits a slight asymmetry. This asymmetry lifts the degeneracy of orthogonal flexural modes. For the NW used here, this resulted in a frequency splitting of $6.26$ kHz between the two fundamental flexural modes (see Fig.~\ref{fig:Figure1}d). To implement force microscopy on a sample surface, the NW is kept attached perpendicularly to its growth substrate and is used as a cantilever in the pendulum geometry (see Fig.~\ref{fig:Figure1}a), with the principal NW axis aligned perpendicular to the sample surface. Displacement of the NW is measured through a fiber-based optical interferometer focused onto the NW\cite{Hoegele08}. While scanning a sample surface underneath the NW (at a fixed distance of 70 nm to the NW tip), we monitor the power spectral density of the thermal motion of the modes. The angle of the two modes with respect to the optical axis of the interferometer can be determined from the relative power in the two resonances to be $\theta_0 = \arctan(\frac{f_1}{f_2}\sqrt{\frac{P_1}{P2}})$, with $f_1$ and $f_2$ the frequencies of the two modes, and $P_1$ and $P_2$ their integrated power\cite{Rossi17}. This allows the tracking of the absolute displacement of the modes as well as rotations of the modes, as the NW is scanned over a sample surface. The sample studied here is patterned with metallic surface gate electrodes (see Fig.~\ref{fig:Figure1}a and c), which are used to generate an electric field of known spatial profile. The experiment takes place inside a helium bath cryostat, at a temperature of 4 K and a vacuum of $<10^{-6}$ mbar. A more detailed description of the experimental setup and techniques can be found in Rossi et al.\cite{Rossi17}.

\noindent The gate electrodes feature a variety of edges with different orientations in the sample plane, which makes it possible to confirm the direction of motion of the two modes\cite{Rossi17}. Here, we restrict ourselves to two opposing gates (see Fig.~\ref{fig:Figure1}c). We apply a voltage bias over the two gates, which results in an electric field with a simulated spatial profile as shown in Fig.~\ref{fig:Figure1}c. There are two dominant types of interaction between this electric field and the NW. First, the presence of a small excess charge $q$ on the NW creates a tip-sample force $\mathbf{F_q}=q\mathbf{E}$. Second, the polarizability of the NW results in a tip-sample force $\mathbf{F_p} = -\mathbf{\nabla} (\alpha|\mathbf{E}|^2)$, with $\alpha$ the effective polarizability of the NW\cite{Rieger12}. We take both forces into account in the following. The spatial derivatives of these vector force fields modify the dynamics of the NW, as detailed further in the next section.

\section*{Vectorial Force Microscopy: Equations of motion}
\noindent The dynamics of the two-mode system can be modeled as that of two coupled, damped and driven harmonic oscillators\cite{Frimmer14}. Forces acting on the NW modify the corresponding equations of motion. Specifically, spatial force derivatives act on the mode frequencies, in a way similar to an increase or decrease of the two spring constants. For small NW oscillation amplitudes, we can approximate the force field acting on the NW modes $i,j\in \{1,2\}$ around displacement $r=0$ as $F_i \approx F_i(0)+r_j\partial F_i/\partial r_j|_{0}$. The equation of motion can now be written in vectorial form as:
\begin{equation}
\Bigg [\frac{d^2}{dt^2} + \frac{\mathbf{\Gamma}}{m}\frac{d}{dt} + \mathbf{\Omega_{0}^{2}}\Bigg ]\begin{pmatrix}r_1\\r_2\end{pmatrix} - \frac{1}{m} \begin{pmatrix}F_{11} & F_{21}\\F_{12} & F_{22}\\ \end{pmatrix} \begin{pmatrix}r_1\\r_2\end{pmatrix} = \frac{1}{m} \begin{pmatrix}F_1(0) + F_{Th,1}(0)\\F_2(0) + F_{Th,2}(0)\end{pmatrix}
\label{EQ:EOM}
\end{equation}
Here we use the dissipation and unperturbed frequency matrices $\mathbf{\Gamma} \equiv \left(\begin{smallmatrix} \Gamma_1 & 0 \\ 0 & \Gamma_2 \end{smallmatrix} \right)$ and $\mathbf{\Omega_{0}^{2}} \equiv\left(\begin{smallmatrix} \Omega_{01}^{2} & 0 \\ 0 & \Omega_{02}^{2} \end{smallmatrix} \right)$. Furthermore, $m$ is the effective mass (equal for the two modes), and $r_i$ are the mode displacements. $F_{Th,i}$ and $F_i$ represent Langevin and other external forces acting on each of the two modes, respectively. The matrix in the last term on the left-hand-side of Eq.~\ref{EQ:EOM} contains the four spatial force derivatives $F_{ij} \equiv \frac{\partial F_i}{\partial r_j}|_{0}$. Of these, the diagonal elements $F_{ii}$ modify the frequency of mode $i$, while the off-diagonal, or shear, components $F_{ij}$ additionally couple the two modes. This coupling results in hybridization of the modes into two new eigenmodes, which are rotated with respect to the original eigenmodes. In case of conservative forces such as those generated by the electric field under investigation here, $F_{12} = F_{21}$, and orthogonality between the modes is maintained. Explicitly including a dependence on the voltage $V$, the eigenvalues of the hybridized modes obtained by diagonalizing the $F_{ij}$ matrix can be written as (for $\Gamma_i/m << \Omega_{0,i}$):
\begin{multline}
  \Omega_{e1,e2} = \frac{1}{\sqrt{2}} \bigg [\Omega_{01}^2 + \Omega_{02}^2 - \frac{F_{11}(V)}{m} - \frac{F_{22}(V)}{m} \\ \pm \sqrt{\Big(-\Omega_{01}^2 +\Omega_{02}^2 + \frac{F_{11}(V)}{m} - \frac{F_{22}(V)}{m}\Big )^2 + \frac{4 F_{12}(V)^2}{m^2}}  \bigg ]^{1/2}.\label{Eq:newEVs}
\end{multline}

\section*{Avoided crossings}
\begin{figure*}
\includegraphics[width=0.85\textwidth]{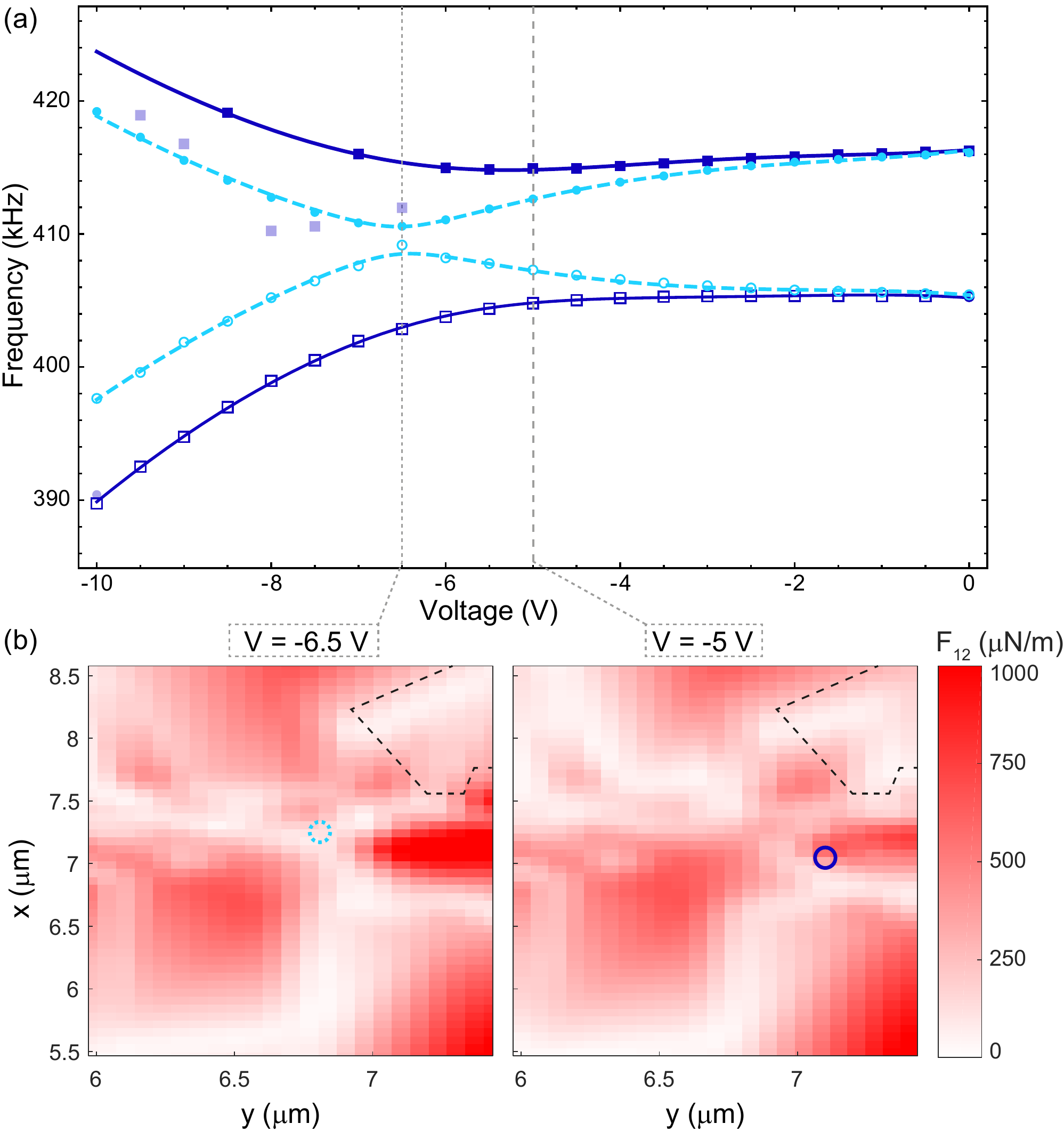}
\caption{(a) Mode frequencies as a function of voltage applied over the two gates. Data points represented by light-blue circles are taken at the position indicated by the dashed circle drawn in the left panel of (b), while the dark blue squares corresponds to data taken at the position of the solid circle drawn in the right panel. The curves are fits of Eq.~\ref{Eq:newEVs} to the data. Grey points are data points not used in the fits. Both sets of data show avoided crossings, with different splittings due to different coupling strengths. (b) Spatial maps of the measured coupling term $F_{12}$ for $V=-6.5$V (left) and $V=-5$V (right). The dashed black line indicates the position of of the nearest gate electrode.}
\label{fig:Figure2}
\end{figure*}

\noindent A signature of strong coupling is formed by mode energies exhibiting avoided crossing under detuning, with a splitting larger than the sum of the linewidths of the individual resonances\cite{Novotny10}. We search for such avoided crossings by measuring the spatial profile of the $F_{ij}$ matrix. We obtain spectra similar to those shown in Fig.~\ref{fig:Figure1}d, for a set of points $(x, y, V)$. From this, we determine the mode frequencies $\Omega_{e,i}/2\pi$ for each spatial point $(x, y)$ as a function of $V$. In Figure \ref{fig:Figure2}a, $\Omega_{e,i}(V)/2\pi$ are plotted for two exemplary positions in the electric field generated by the gate electrodes. It can be seen that the mode frequencies shift under the influence of $V$, and can be made to move away or towards each other. The frequency shifts plotted in Fig.~\ref{fig:Figure2}a indeed exhibit avoided crossings, indicating that the coupling gradient $F_{12}$ is non-zero. At the center of an avoided crossing, at voltage $V_c$, the eigenmodes can be written as $|-\rangle = (1, -1)$ and $|+\rangle = (1,1)$ in the basis of the original modes. The size of the avoided crossing can be approximated as $\Delta\Omega(V_c) = \Omega_+ - \Omega_- \approx F_{12}(V_c)/m\Omega_0$, where $\Omega_0 \equiv \sqrt{\Omega_{01}^2 - \frac{F_{11}(V_c)}{m}} = \sqrt{\Omega_{02}^2 - \frac{F_{22}(V_c)}{m}}$. We see therefore that the splitting is proportional to the coupling force gradient. By moving along the edge of a gate electrode, we observe that we can controllably set $F_{12}$ and observe avoided crossings with correspondingly different gap sizes.

\begin{figure*}
\includegraphics[width=1\textwidth]{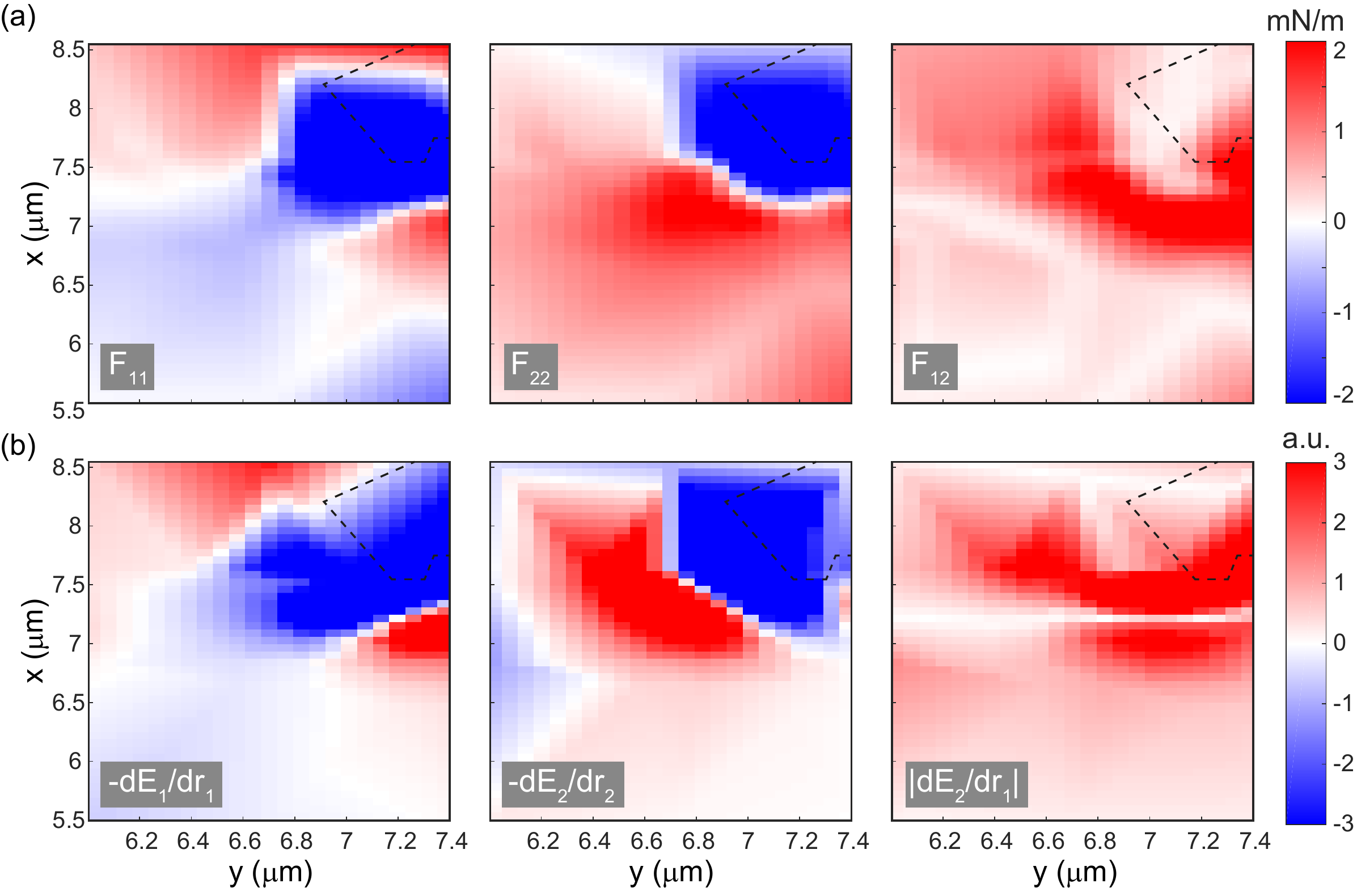}
\caption{(a) Spatial maps of the experimentally determined force derivatives $F_{11}$ (left), $F_{22}$ (middle), and $F_{12}$ (right). (b) Simulated derivatives of the electric field $-dE_1/dr_1$ (left), $-dE_2/dr_2$ (middle), and $|dE_2/dr_1|$ (right). For all plots $V$=-10V.}
\label{fig:Figure3}
\end{figure*}

\noindent By fitting Eq.~\ref{Eq:newEVs} to $\Omega_{e1}(x,y,V)$ and $\Omega_{e2}(x,y,V)$, we directly obtain values for $F_{11}$, $F_{22}$, and $F_{12}$ as a function of $(x,y,V)$. Fig.~\ref{fig:Figure2}b shows two spatial maps of $F_{12}$, for $V_c$ of the avoided crossings displayed in Fig.~\ref{fig:Figure2}a. Consistent with expectations, the larger avoided crossing occurs at a position where $F_{12}$ is larger. Similarly, we obtain spatial maps of $F_{11}$, $F_{22}$, and $F_{12}$ for a range of voltages $V$. In Figure \ref{fig:Figure3}, we compare two of these spatial maps with predicted values based on finite-element modeling of the electrostatic field generated by the gate electrodes (COMSOL). In Fig.~\ref{fig:Figure3}a, measured values of $F_{11}(x,y)$ (left), $F_{22}(x,y)$ (center), and $F_{12}(x,y)$ (right) are shown for $V=-10V$. Fig.~\ref{fig:Figure3}b displays the corresponding spatial derivatives of the simulated electric field. Here we plot only the first order derivatives of the electric field, since these are found to give the dominant contribution. The good agreement between theory and experiment indicates that such modeling can be used to qualitatively predict the values of $F_{ij}$ and engineer the properties of the two-mode system, including the coupling strength. Note that at voltages close to zero, the force derivatives also contain significant contributions from forces other than those arising from the applied electric field, such as Van der Waals forces.

\section*{Rabi Oscillations}
In coupled two-level systems, energy can be coherently exchanged between the two levels through driven Rabi oscillations. Rabi oscillations in the coupled two-mode system can be induced through periodic modulation of the frequency detuning\cite{Frimmer14}. When the modulation frequency $\omega_{d}$ is close to the frequency difference $\Delta\Omega$, a coherent oscillation between the populations of the two modes will take place. Here, modulation of the detuning is generated by applying a drive voltage $V_{ac}(t) = V_{d}\cos(\omega_{d} t)$ over the two gates. 

\noindent In a classical picture, Rabi oscillations between the populations $|a_{\pm}|^2$ of the two modes, including decay, can be described as\cite{Frimmer14}
\begin{equation}
|a_-(t)|^2 = \sin^2\bigg(\frac{\Omega_R t}{2}\bigg)e^{-\Gamma t},~~ 
|a_+(t)|^2 = \cos^2\bigg(\frac{\Omega_R t}{2}\bigg)e^{-\Gamma t}
\label{eq:rabi}
\end{equation}

\noindent Here, initialization is assumed to give $|a_-(0)|^2 = 0$, $|a_+(0)|^2 = 1$. Furthermore, $\Omega_R = \sqrt{A^2 + \delta_{d}^2}$ is the generalized Rabi oscillation frequency, with the amplitude $A$ set by $V_{d}$, and $\delta_d = \omega_d - \Delta\Omega$. We use a single decay constant $\Gamma$ for both modes, since for the two hybridized modes (at the center of the avoided crossing), population decay rates should be equal. $\Gamma$ corresponds to the time constant of evolution towards $|a_-(0)|^2 = 0$, $|a_+(0)|^2 = 0$. In a Bloch sphere representation of the populations, this corresponds to a shrinking of the state vector.

\begin{figure*}
\includegraphics[width=1\textwidth]{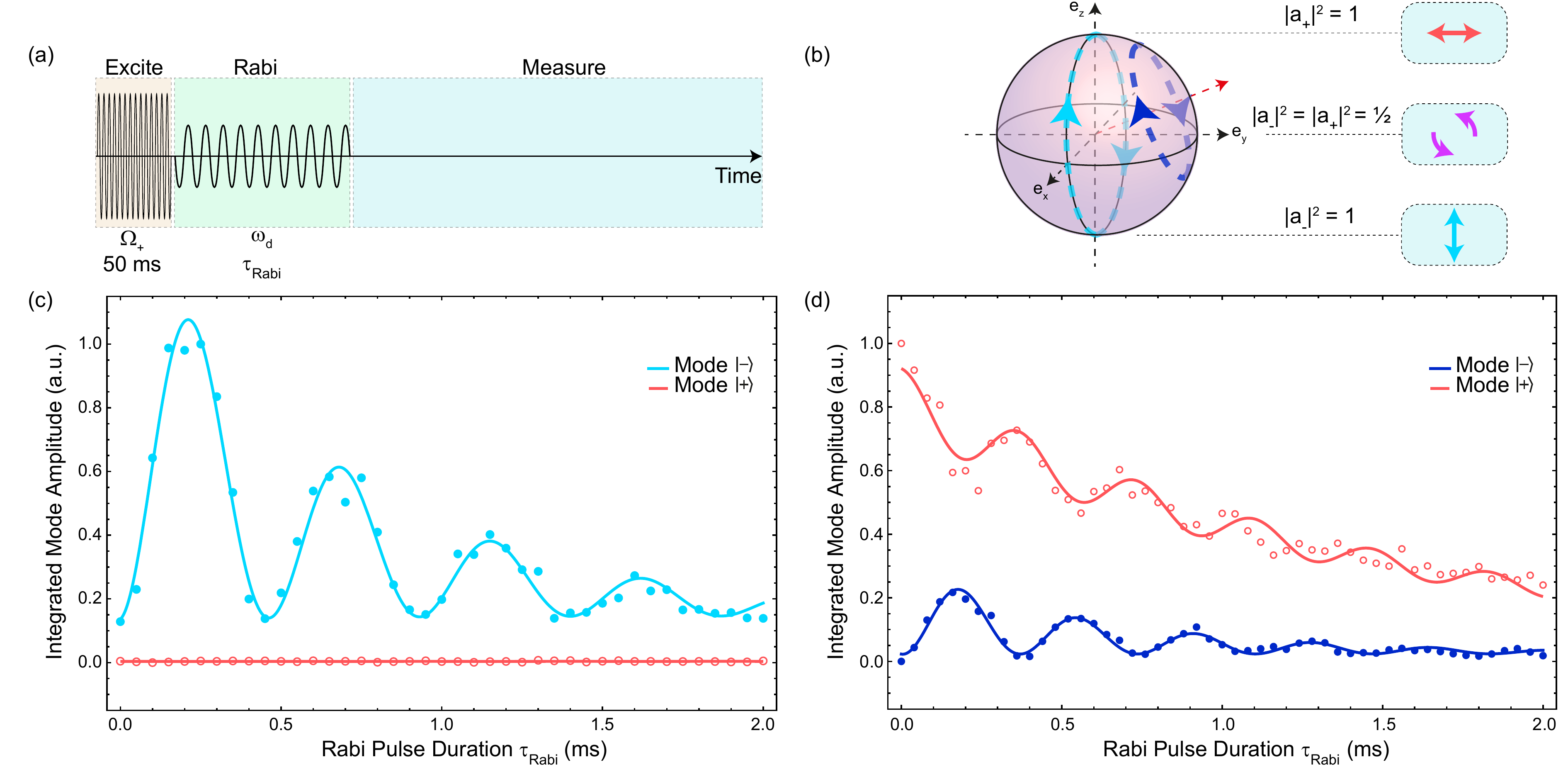}
\caption{
(a) Schematic of Rabi pulsing scheme. (b) Bloch sphere representation of Rabi oscillations. Light blue (vertical) arrow indicates Rabi oscillations for $\delta_d=0$, dark blue (tilted) arrow indicates Rabi oscillations for $\delta_d\neq 0$. Boxed arrows on the right side schematically indicate NW in the plane for different population distributions. (c) Power spectral densities of both modes as a function of Rabi pulse duration. Points indicate measured data, solid lines are fits to the data. The fit for mode $|-\rangle$ returns a Rabi frequency of 1.9 kHz ($\Delta\Omega/2\pi = 10.9$kHz). (d) Similar to (c), but now for $\delta_d \neq 0$. Fits yield Rabi frequencies of 2.8 kHz (mode $|-\rangle$) and 3.0 kHz (mode $|+\rangle$)  ($\Delta\Omega/2\pi = 12.1$kHz). Note that the offset in the measured Rabi oscillations is due to thermal excitation of the mode.
}
\label{fig:Figure4}
\end{figure*}

\noindent In order to excite and detect Rabi oscillations in our system, we implement the following measurement protocol (see Fig.~\ref{fig:Figure4}a). First, we excite mode $|+\rangle$ by applying a drive pulse with frequency $\Omega_+/2\pi$ for 50 ms. Next, a pulse of varying duration $\tau_{Rabi}$ with frequency $\omega_{d}/2\pi$ is applied to drive the Rabi oscillations. During this pulse, a coherent exchange of energy takes place between the two modes, resulting in a change of the two populations. Finally, the populations of the two modes are read out by measuring the power spectral density of each mode. We average over 50 such cycles.

\noindent Figure \ref{fig:Figure4}c shows Rabi oscillations for $\delta_d=0$. Here the modes are aligned approximately parallel (mode $|-\rangle$) and perpendicular (mode $|+\rangle$) to the optical axis of the interferometer. Consequently, the signal-to-ratio of the displacement measurement of mode $|+\rangle$ is very small, and Rabi oscillations are hard to detect. Instead, for a small but finite $\delta_d$, the modes are rotated with respect to the optical axis, and Rabi oscillations can be detected for both modes (see Fig.~\ref{fig:Figure4}d). In this case, the Rabi oscillations result in only a partial transfer of populations, as can be understood from the schematic depiction of the Bloch sphere in Fig.~\ref{fig:Figure4}b. For $\delta_d=0$, the Rabi frequency depends linearly on the drive amplitude $A$, where $A$ is a quadratic function of $V_d$. We observe Rabi frequencies in the kHz-regime, with a maximum of 4.5 kHz, as determined from fitting Eq.~\ref{eq:rabi} to the data. The same fits return values for the relaxation time $T_1$ between $0.6$ and $1.5$ ms, in agreement with independent ringdown measurements. 

\noindent Interestingly, our driven two-mode system is close to the regime of strong driving, with the Rabi frequency $\Omega_R$ approaching the transition frequency $\Delta\Omega$. We observe a maximum of roughly $\Omega_R = \Delta\Omega/3$. When a two-level system is strongly driven, its dynamics is not simply sinusoidal anymore, but can become anharmonic and nonlinear. In other systems, ripples on top of sinusoidal oscillations have been observed\cite{Spreeuw90}, as well as oscillations that are faster than expected from the Rabi model\cite{Fuchs09,Scheuer14,Rama17}. Such effects may explain irregularities in our measurements. Furthermore, note that in our experiments we were limited by our electronics allowing only a limited amplitude $A$ of the Rabi drive, and higher Rabi frequencies should be feasible. This would enable further studies of the strong driving regime in the case of a classical two-mode system.

\section*{Conclusions}
Quantum control techniques such as dynamical decoupling rely on coherent dynamics of two-level systems. Coherent dynamics of classical two-mode mechanical oscillators, such as those demonstrated here, open the way to use similar control techniques for sensing applications. A recent review\cite{Sansa16} of experimental work has shown that frequency fluctuations of nanomechanical oscillators are often much higher than the thermal limit. Dynamical decoupling pulse sequences offer a way to reduce frequency fluctuations, potentially down to the thermal limit. This method would be applicable in a wide range of mechanical oscillators, requiring only the presence of two strongly coupled modes and the ability to drive several Rabi oscillations within the relaxation time.

\newpage

\newpage

\newpage

\newpage

\end{document}